\documentclass[twocolumn,showpacs,preprintnumbers,superscriptaddress,amsmath,floatfix,amssymb,prd]{revtex4}
\usepackage[colorlinks=true]{hyperref}
\usepackage{graphicx}

\newcommand{\comment}[1]{}

\newcommand{\lr}[1]{ \left( #1 \right) }
\newcommand{\lrs}[1]{ \left[ #1 \right] }

\newcommand{\vev}[1]{ \langle \, #1 \, \rangle }

\newcommand{\tr}{ {\rm Tr} \, }

\newcommand{\ket}[1]{ \, | #1 \rangle }
\newcommand{\bra}[1]{ \langle #1 | \, }

\newcommand{\expa}[1]{ \exp{\left( #1 \right)} }

\newcommand{\logo}{\\ \vskip -18mm
\leftline{\includegraphics[scale=0.3,clip=false]{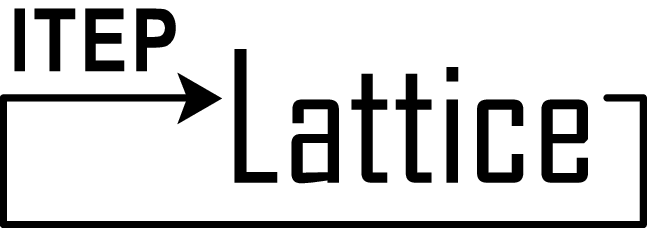}} \vskip 10mm}

\begin{document}
\sloppy
\preprint{ITEP-LAT/2008-07}

\title{Numerical study of entanglement entropy in $SU\lr{2}$ lattice gauge theory \logo}
\author{P. V. Buividovich}
\email{buividovich@tut.by}
\affiliation{JIPNR, National Academy of Science, 220109 Belarus, Minsk, Acad. Krasin str. 99}
\affiliation{ITEP, 117218 Russia, Moscow, B. Cheremushkinskaya str. 25}
\author{M. I. Polikarpov}
\email{polykarp@itep.ru}
\affiliation{ITEP, 117218 Russia, Moscow, B. Cheremushkinskaya str. 25}
\date{April 15, 2008}
\begin{abstract}
 The entropy of entanglement between a three-dimensional slab of thickness $l$ and its complement is studied numerically for four-dimensional $SU\lr{2}$ lattice gauge theory. We find a signature of a nonanalytic behavior of the entanglement entropy, which was predicted recently for large $N_{c}$ confining gauge theories in the framework of AdS/CFT correspondence. The derivative of the entanglement entropy over $l$ is likely to have a discontinuity at some $l = l_{c}$. It is argued that such behavior persists even at finite temperatures, probably turning into a sort of crossover for temperatures larger than the temperature of the deconfinement phase transition. We also confirm that the entanglement entropy contains quadratically divergent $l$-independent term, and that the nondivergent terms behave as $l^{-2}$ at small distances.
\end{abstract}
\maketitle

\section{Introduction}
\label{sec:intro}

 The application of the concepts of quantum information theory to quantum field theories in continuous space-times and on the lattices has led recently to many important advances in our understanding of their quantum behavior \cite{Kitaev:03:1, Calabrese:06:1, Osborne:02:1, Calabrese:04:1, Kitaev:06:1, Osborne:02:2}. One such concept is the quantum entanglement of states of systems with many degrees of freedom, which turned out to be a very useful model-independent characteristic of the structure of the ground state of quantum fields. In particular, for quantum fields on the lattice in the vicinity of a quantum phase transition (i.e. a phase transition which occurs at zero temperature when some parameters of the system, such as the coupling constants, are varied) the ground state is a strongly entangled superposition of the states of all elementary lattice degrees of freedom (such as spins in Heisenberg model, or link variables in lattice gauge theory), and different phases of lattice theories can be characterized by different patterns of entanglement \cite{Kitaev:03:1, Calabrese:06:1, Osborne:02:1, Calabrese:04:1, Kitaev:06:1, Osborne:02:2}. Quantum entanglement thus appears to be an adequate concept for the description of the emergence of collective degrees of freedom in quantum field theories \cite{Kitaev:03:1, Calabrese:06:1, Osborne:02:1, Calabrese:04:1, Kitaev:06:1}.

 A commonly used measure of quantum entanglement of the ground state of quantum fields in $\lr{D-1} + 1$-dimensional space-time is the entropy of entanglement $S\lrs{A}$ between some $\lr{D-1}$-dimensional region $A$ and its $\lr{D-1}$-dimensional complement $B$, which characterizes the amount of information shared between $A$ and $B$ \cite{Bernstein:96:1, Calabrese:06:1, Calabrese:04:1}. Entanglement entropy is defined as the usual von Neumann entropy for the reduced density matrix $\hat{\rho}_{A}$ associated with the region $A$:
\begin{eqnarray}
\label{ent_def}
S\lrs{A} = - \tr_{A} \lr{ \hat{\rho}_{A} \ln \hat{\rho}_{A} }
\end{eqnarray}
The reduced density matrix is obtained from the density matrix of the ground state of the theory, $\hat{\rho}_{AB} = \ket{0} \bra{0}$, by tracing over all degrees of freedom which are localized outside of $A$, i.e. within $B$ \cite{Bernstein:96:1, Calabrese:06:1}:
\begin{eqnarray}
\label{dm_def}
\hat{\rho}_{A} = \tr_{B} \hat{\rho}_{AB} = \tr_{B} \ket{0}\bra{0}
\end{eqnarray}
 This density matrix describes the state of quantum fields as seen by an observer who can only perform measurements within $A$. In other words, the region $B$ is inaccessible for such an observer, as if it was separated from $A$ by a sort of event horizon. Thus the entanglement entropy is in some sense similar to the entropy of black holes \cite{Srednicki:93:1, Fursaev:06:1}.

 As was demonstrated by Bekenstein and Hawking, the entropy of a black hole is proportional to the area of its horizon \cite{Hawking:73:1, Bekenstein:75:1}. This fact has recently received an interesting development in the works \cite{Ryu:06:1, Ryu:06:2, Takayanagi:06:1, Klebanov:07:1}, where this entropic ``area law'' was conjectured to hold also for the entanglement entropy of field theories which are dual (in the sense of Maldacena, or gauge-gravity, duality \cite{Maldacena:97:1}) to supergravity on AdS spaces (or their thermal modifications). More precisely, the original conjecture of \cite{Ryu:06:1, Ryu:06:2} is that for $\lr{D-1} + 1$-dimensional conformal field theories living on the boundary of $\lr{D+1}$-dimensional AdS space, the entropy of entanglement between the region $A$ and its complement is proportional to the minimal area of hypersurface in AdS space which is spanned on the boundary of $A$. This was explicitly demonstrated for two-dimensional conformal field theories living on the boundary of $AdS_{3}$, for which the entanglement entropy is proportional to the minimal length of a line which joins the ends of $A$ in the bulk of this $AdS_{3}$ (for field theories in $1+1$ space-time dimensions $A$ is just a line). For higher-dimensional conformal field theories which have supergravity duals the entanglement entropy is not known exactly, thus the conjecture of \cite{Ryu:06:1, Ryu:06:2} cannot be directly verified. Nevertheless, it was found in \cite{Ryu:06:1, Ryu:06:2} that the ``area law'' prescription gives at least a reasonable approximation both for zero and nonzero temperatures also in these cases.

 To be more specific, the results of \cite{Takayanagi:06:1, Klebanov:07:1, Ryu:06:1, Ryu:06:2} are obtained for the case when $A$ is a slab of thickness $l$ in $\lr{D-1}$-dimensional space, and the entanglement entropy is normalized per unit area of the boundary $\partial A$ of $A$. A typical behavior of entanglement entropy for field theories in $\lr{D-1} + 1$-dimensional space-time is \cite{Takayanagi:06:1, Klebanov:07:1, Ryu:06:1, Ryu:06:2}:
\begin{eqnarray}
\label{ent_asymp}
\frac{1}{|\partial A|} \, S\lr{l} = \frac{1}{|\partial A|} \, S_{UV} + \frac{1}{|\partial A|} \, S_{f}\lr{l}
= \nonumber \\ =
\kappa \, \Lambda_{UV}^{D-2} - l^{2 - D} f\lr{l}
\end{eqnarray}
where $\kappa$ is some constant which is different for different theories, $\Lambda_{UV}$ is the UV cutoff scale and $f\lr{l}$ is some function which is finite as $l \rightarrow 0$. The $l^{2 - D}$ behavior of the UV finite term in (\ref{ent_asymp}) can be fixed by the requirement that the entropy does not diverge at $l \rightarrow 0$, which is reasonable for a local field theory. Indeed, the ultraviolet divergence in (\ref{ent_asymp}) can only be cancelled if $S_{f}\lr{l}$ tends to $ - \kappa \, l^{2 - D}$ at small distances of order of $\Lambda_{UV}^{-1}$. This means that the entanglement entropy associated with a set of points of zero measure is finite. If the entanglement entropy is calculated as the minimal area of hypersurface with boundary $\partial A$ in the bulk geometry, the ultraviolet divergence and the $l^{2 - D}$ behavior at small distances are automatically reproduced due to the singularity of the metric near the boundary of AdS$_{D+2}$. Thus the nontrivial information about the entanglement due to field interactions is contained in the function $f\lr{l}$. In terms of the dual geometries the function $f\lr{l}$ is a probe of the nonsingular structure of the bulk space. For the sake of brevity we will restrict further discussion to the case $D = 4$. Following \cite{Takayanagi:06:1}, it is convenient to introduce the so-called entropic $C$-function as follows:
\begin{eqnarray}
\label{entropic_C_function}
C\lr{l} = \frac{l^{3}}{|\partial A|} \, \frac{\partial}{\partial l} \, S\lr{l}
 = 2 \, f\lr{l} - l \, \frac{\partial}{\partial l} \, f\lr{l}
\end{eqnarray}
Since the divergent part of the entropy (\ref{ent_asymp}) does not depend on $l$, ultraviolet divergences cancel in (\ref{entropic_C_function}), and the entropic $C$-function is UV finite. The factor $l^{3}$ ensures that $C\lr{l}$ is also finite at $l \rightarrow 0$. For conformal field theories, this function is proportional to the central charge and thus counts the number of effective degrees of freedom of the theory, that is, the number of distinct field species weighted with their central charges (such as $1$ for a free boson, $1/2$ for a free fermion etc.). An important property of the entropic $C$-function, which follows from Zamolodchikov $C$-theorem \cite{Zamolodchikov:86:1}, is that it should decrease as the energy scale degreases, i.e. as its argument $l$ grows.

 Simplicity and elegance of the ``area law'' for the entanglement entropy suggest that it has a fundamental nature and can be applied also to nonconformal field theories which have dual supergravity descriptions. This was done in \cite{Takayanagi:06:1, Klebanov:07:1}, where the entanglement entropy of some confining large $N_{c}$ gauge theories was studied basing on the conjecture of \cite{Ryu:06:1, Ryu:06:2}. These theories are dual to supergravity on certain SUSY-breaking five-dimensional geometries \cite{Witten:98:1, Klebanov:00:2}. One of the most interesting results of \cite{Takayanagi:06:1, Klebanov:07:1} is that if the entropic ``area law'' is assumed to hold also for such theories, their entanglement entropy should be nonanalytic in $l$. This nonanalytic behavior is a consequence of the existence of two distinct minimal hypersurfaces in five-dimensional bulk geometry, a connected one and a disconnected one \cite{Takayanagi:06:1, Klebanov:07:1}. For small regions $A$ the area of the connected hypersurface is less than that of the disconnected one, and it is this area which determines the entanglement entropy, while for sufficiently large $A$ the situation is reversed. At some $l_{c}$ the areas of these two hypersurfaces become equal, and as a result the derivative of the entanglement entropy over $l$ and the entropic $C$-function are discontinuous at $l = l_{c}$. For example, for gauge theories which are dual to $D3$ or $D4$ branes on a circle \cite{Klebanov:07:1}, $C\lr{l}$ is proportional to the step function: $C\lr{l} \sim N_{c}^{2} \theta\lr{l_{c} - l}$. This means that the effective number of degrees of freedom rapidly changes from being of order of $N_{c}^{2}$ to $N_{c}^{0}$ as the region $A$ in (\ref{ent_def}) becomes smaller (in the large $N_{c}$ limit $N_{c}^{0}$ is just zero as compared with $N_{c}^{2}$). It was conjectured in \cite{Klebanov:07:1} that this nonanalytic behavior is similar to the confinement-deconfinement phase transition at finite temperatures. Namely, when the size of the region $A$ is sufficiently small and the corresponding energy scale is sufficiently high, we are in the domain of asymptotic freedom and the number of particles which contribute to the entropy (\ref{ent_def}) is proportional to $N_{c}^{2}$. As the region $A$ becomes larger and the infrared effects become strong, coloured states are confined and we are left with a theory of colourless glueballs and hadrons, for which the number of states and the entropy are much smaller. It was argued in \cite{Klebanov:07:1} that such behavior of the entanglement entropy is generic for confining large-$N_{c}$ gauge theories because of the Hagedorn-like structure of their spectrum.

 Thus the behavior of entanglement entropy of confining gauge theories can be an interesting new test of Maldacena duality between string theories on $\lr{D+1}$-dimensional curved spaces and $D$-dimensional gauge theories which live on their boundary. It can be also used as an alternative probe which indicates the emergence of the colourless degrees of freedom in the low-energy limit of gauge theories. The results of \cite{Takayanagi:06:1, Klebanov:07:1} imply that colourless degrees of freedom always emerge instantly, so that in some sense gluons and glueballs cannot coexist at one energy scale. For finite $N_{c}$ the arguments of \cite{Takayanagi:06:1, Klebanov:07:1} remain valid at the qualitative level, however, it is not clear how the nonanalytic behavior may be modified in this case, and different phenomenological descriptions of confining gauge theories give different predictions. For instance, in the AdS space with hard wall \cite{Polchinski:02:1}, which is one of the most popular phenomenological stringy models of confinement, there is only one minimal hypersurface for all $l$ \cite{Klebanov:07:1}, and according to the conjecture of \cite{Ryu:06:1, Ryu:06:2, Takayanagi:06:1, Klebanov:07:1} the entanglement entropy should be analytic in $l$. On the other hand, if the entanglement entropy is calculated using the approximate Migdal-Kadanoff decimations in lattice gauge theory, it appears to be non-analytic \cite{Velytsky:08:1}.

 Unfortunately, up to now the infrared dynamics of confining gauge theories is not understood from the first principles, and no method is known which can be used to calculate explicitly the entanglement entropy of confining gauge theories for sufficiently large regions $A$. The aim of the present work is to investigate the entanglement entropy (\ref{ent_def}) in pure $SU\lr{2}$ gauge theory numerically, using lattice Monte-Carlo simulations. The same geometry of $A$ as in \cite{Takayanagi:06:1, Ryu:06:1, Ryu:06:2, Klebanov:07:1, Velytsky:08:1} is used, namely, $A$ is a slab of three-dimensional space which has thickness $l$ in one direction and which extends maximally in all other directions, and the entropy is normalized per unit area of the boundary of $A$. Simulation results suggest that the derivative of the entanglement entropy over $l$ is indeed discontinuous at some $l = l_{c}$. In addition, we present some indications that this nonanalytic behavior may persist even at finite temperatures above $T_{c}$, probably turning into a crossover line in the $l \; - \; T$ plane. We also confirm that the entanglement entropy diverges quadratically and that at small distances $S_{f}\lr{l}$ indeed behaves as $S_{f}\lr{l} \sim - \, l^{-2}$.

 The structure of the paper is the following. In Section \ref{sec:ent_path_int} we discuss the path-integral representation of the entanglement entropy, which is suitable for Monte-Carlo lattice simulations, and introduce the spaces of special topology on which these simulations should be performed. In Section \ref{sec:act_dist} we present the results of our measurements of the action density distributions on lattices with such special topology and confirm that the divergent term in the entanglement entropy scales as $\Lambda_{UV}^{2}$ and does not depend on $l$. In Section \ref{sec:ent_growth} the dependence of the nondivergent part of the entanglement entropy on $l$ is investigated and the discontinuity of the derivative $\frac{\partial}{\partial l} \, S_{f}\lr{l}$ is detected. Nonanalytic behavior of the entanglement entropy with respect to $l$ at $T > T_{c}$ is discussed and studied numerically in Section \ref{sec:lt_phase_tr}. Finally, Section \ref{sec:conclusions} contains some concluding remarks and the discussion of the obtained results. Some technical details of our calculations are relegated to the Appendices.

\section{Measurement procedure}
\label{sec:ent_path_int}

 In this Section we present some important details of the numerical procedures which we have used to measure the entanglement entropy. Highly technical details are relegated to Appendices \ref{sec:path_int_der} and \ref{sec:mc_fe}.

 Entanglement entropy can not be expressed as an expectation value of some field operator, thus theoretical and especially numerical calculations of such entropy in field theories are far from being straightforward. For example, most investigations of quantum entanglement of the ground states of lattice spin systems were carried out using rather special transfer-matrix methods, which are applicable only to some particular models \cite{Kitaev:03:1}. In \cite{Calabrese:04:1} a path-integral representation of the entanglement entropy was derived, which relates the entanglement entropy to the set of partition functions on the spaces of some special topology, namely, the spaces with an integer number of cuts. In two dimensions such spaces can be implemented as multi-sheeted Riemann surfaces, which allows one to find explicitly the entanglement entropy in conformal field theories \cite{Calabrese:04:1, Calabrese:06:1}.

\begin{figure}[ht]
  \includegraphics[width=7cm]{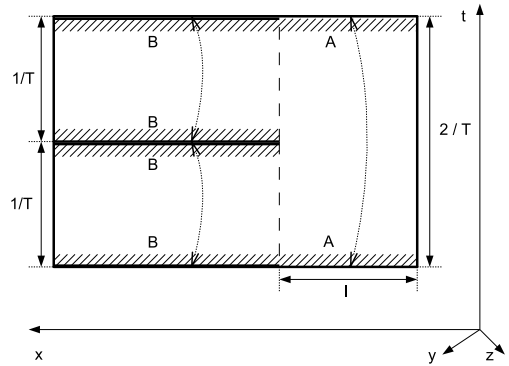}\\
  \caption{The topology of space on which the free energy $F\lrs{A, s, T}$ in (\ref{ent_vs_fe}) is calculated, an example for $s = 2$. Dashed lines with arrows denote identification of cut sides, i.e. periodic boundary conditions in time direction.}
  \label{fig:cutted_box}
\end{figure}

 For higher-dimensional spaces, it is also relatively easy to implement the spaces of required topology, at least on the lattice. For the purposes of lattice simulations it is convenient to represent the result of \cite{Calabrese:04:1} in the following form:
\begin{eqnarray}
\label{ent_vs_fe}
S\lrs{A} = \lim \limits_{T \rightarrow 0} \lr{
\lim \limits_{s \rightarrow 1} \, \frac{\partial}{\partial s} \, F\lrs{A, s, T}
- F\lr{T}
}
\end{eqnarray}
where $F\lr{T} = - \ln \mathcal{Z}\lr{T}$ is the free energy of the theory at the temperature $T$. $\mathcal{Z}\lrs{A, s, T}$ and $F\lrs{A, s, T} = - \ln \mathcal{Z}\lrs{A, s, T}$ are the partition function and the free energy obtained by integrating over all fields with the following boundary conditions \cite{Calabrese:04:1, Calabrese:06:1}: the fields are periodic in time with the period $s T^{-1}$ if the spatial coordinates lie within $A$, otherwise (i.e. for spatial coordinates within $B$) the period is $T^{-1}$. Such topology of space is schematically shown on Fig. \ref{fig:cutted_box} for $s = 2$. Thus $F\lrs{A, s, T}$ is the free energy of the theory on the space with $s$ cuts \cite{Calabrese:04:1, Calabrese:06:1}. In four-dimensional space-time these cuts are the three-volumes $B$ at the time slices $t = k T$, $k = 0, 1, \ldots, s$, which have a common boundary $\partial A = \partial B$. At this boundary the branching points of the cuts are situated. The expression (\ref{ent_vs_fe}) is derived in Appendix \ref{sec:path_int_der}, where the original derivation of \cite{Calabrese:04:1} is briefly revised.

 Since free energies can not be directly measured in lattice simulations, one usually looks at the average plaquette action, which is related to the derivative of the free energy over the coupling constant $\beta = \frac{1}{4 g_{s}^{2}}$ of $SU\lr{2}$ lattice gauge theory:
\begin{eqnarray}
\label{free_enrg_vs_plaq}
\sum \limits_{p} \, \vev{ \beta \lr{1 - \frac{1}{2}\,\tr g_{p}} } =
\frac{\partial}{\partial \ln{\beta}} \, F\lr{\beta}
\end{eqnarray}
where summation goes over all lattice plaquettes. The average plaquette action and the free energy have the same power-like ultraviolet divergences, which follows from the fact that the lattice spacing $a\lr{\beta} = \Lambda_{UV}^{-1}$ depends on $\beta$ as $a\lr{\beta} \sim \beta^{\alpha} \expa{ - \gamma \beta}$, where the coefficients $\alpha$ and $\gamma$ can be obtained from the Gellman-Low $\beta$-function. Thus the derivative in (\ref{free_enrg_vs_plaq}) can only change logarithmic dependence on the UV cutoff scale, which is usually too weak to be observed in lattice simulations. In Section \ref{sec:act_dist} we use the expression (\ref{free_enrg_vs_plaq}) to find the UV divergent part of the entanglement entropy (\ref{ent_asymp}).

 In practice it turns out that the equation (\ref{free_enrg_vs_plaq}) can only be reliably used to find the divergent part of the entanglement entropy, while extremely precise measurements are required to study the behavior of the nondivergent part. For this reason it is more convenient to measure the derivative of the entanglement entropy (\ref{ent_vs_fe}) over $l$, which is related to the free energy as:
\begin{eqnarray}
\label{ent_der_vs_fe_der}
\frac{\partial}{\partial l} \, S\lrs{A} = \frac{\partial}{\partial l} \, S_{f}\lr{l}
 =
\lim \limits_{T \rightarrow 0} \lim \limits_{s \rightarrow 1} \, \frac{\partial}{\partial s} \, \frac{\partial}{\partial l} \, F\lrs{A,s,T}
\end{eqnarray}
The derivative $\frac{\partial}{\partial l} \, F\lrs{A,s,T}$ can be estimated from finite differences of free energies for close values of $l$. Such differences can be directly measured in lattice simulations using the method proposed recently in \cite{Fodor:07:1, Fodor:07:2}. The results of these measurements are presented in Section \ref{sec:ent_growth}, and the method itself is described in Appendix \ref{sec:mc_fe}.

 Thus the expressions (\ref{ent_vs_fe}), (\ref{free_enrg_vs_plaq}) and (\ref{ent_der_vs_fe_der}) allow one to calculate the entanglement entropy by performing lattice simulations on lattices with integer number of cuts. In practice such simulations can be done only for several smallest $s$, after which some numerical interpolation method should be used to find the derivative over $s$. In this work we use the simplest possible way to estimate this derivative. Namely, from lattice simulations on lattices with two cuts we obtain the values of $\lim \limits_{T \rightarrow 0} F\lrs{A,2,T}$ or $\lim \limits_{T \rightarrow 0} \frac{\partial}{\partial l} \, F\lrs{A,2,T}$. Taking into account that by definition $\frac{\partial}{\partial l} \, F\lrs{A,1,T} = 0$ and $F\lrs{A,1,T} = F\lr{T}$, we estimate the derivative over $s$ at $s = 1$ in (\ref{ent_vs_fe}) as the following finite difference:
\begin{eqnarray}
\label{working_def}
\lim \limits_{T \rightarrow 0} \lr{
\lim \limits_{s \rightarrow 1} \, \frac{\partial}{\partial s} \, F\lrs{A, s, T}
- F\lr{T}
}
\approx \nonumber \\ \approx
\lim \limits_{T \rightarrow 0} \lr{F\lrs{A,2,T} - 2 F\lr{T}}
\end{eqnarray}
The results that we obtain using such an estimation are in a rather good agreement with theoretical expectations \cite{Takayanagi:06:1, Klebanov:07:1, Ryu:06:1, Ryu:06:2}, which gives us a hope that the approximation (\ref{working_def}) is not unreasonable. It should be also noted that in fact any of the quantities $P_{A}\lr{s} = \expa{ - F\lrs{A,s,T} + s F\lr{T}} = \tr \hat{\rho}_{A}^{s}$ (see Appendix \ref{sec:path_int_der} for a proof of this equality) for $s > 1$ can be used as a measure of quantum entanglement. In particular, the quantity $P = \expa{ - F\lrs{A,2,T} + 2 F\lr{T}} = \tr \hat{\rho}_{A}^{2}$, although it does not have such a straightforward physical interpretation as the entanglement entropy, is also widely used in quantum information theory and has a special name of the purity of quantum states. For pure states the purity is equal to $1$, for impure states it is less than $1$, and the smaller it is, the more impure is the quantum state described by $\hat{\rho}_{A}$.

\section{Average action density on lattices with cuts}
\label{sec:act_dist}

 In this Section we present the results of our measurements of the average action density on lattices with cuts, and demonstrate that the entanglement entropy contains quadratically divergent $l$-independent term.

 Lattices with $s$ cuts are obtained from $s \, N_{t} \times N_{L}^{3}$ lattices with periodic boundary conditions in all space-like directions by identifying lattice sites with periods $N_{t}$ and $s N_{t}$ in time for spatial coordinates within $A$ and $B$, respectively. These sites are then linked taking into account this periodical identification, and lattice gauge theory is defined on the obtained links. It turns out that lattices with such topology can be entirely mapped on $s N_{t} \times N_{L}^{3}$ hypercubic lattices so that all $s \, D \, N_{t} \times N_{L}^{3}$ links and $s \, \frac{D\lr{D - 1}}{2} \, N_{t} \times N_{L}^{3}$ lattice plaquettes are used. To reduce the anisotropy due to the cuts, in our simulations at each Monte-Carlo iteration the heatbath and the overrelaxation procedures are applied to $2 \, s \, D \, N_{t} \times N_{L}^{3}$ randomly chosen links.

\begin{figure*}[ht]
  \includegraphics[width=6cm]{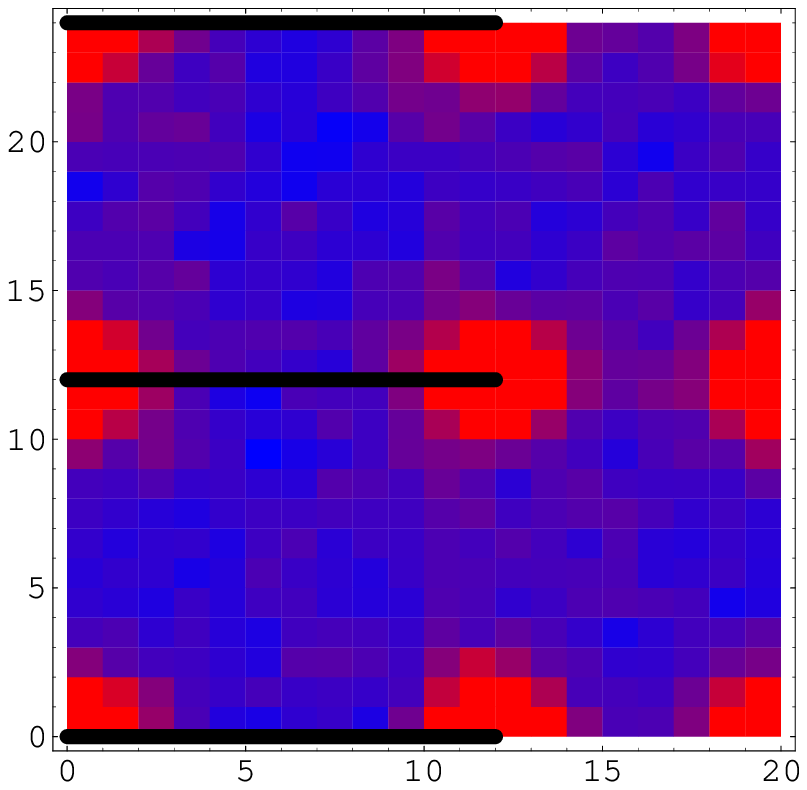}
  \includegraphics[width=6cm]{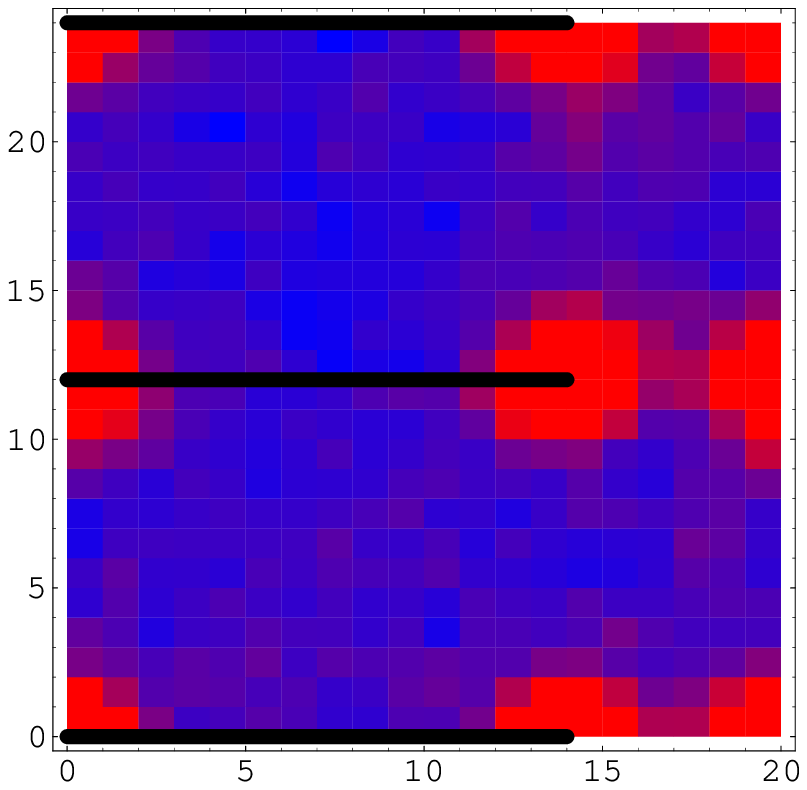}\\
  \includegraphics[width=6cm]{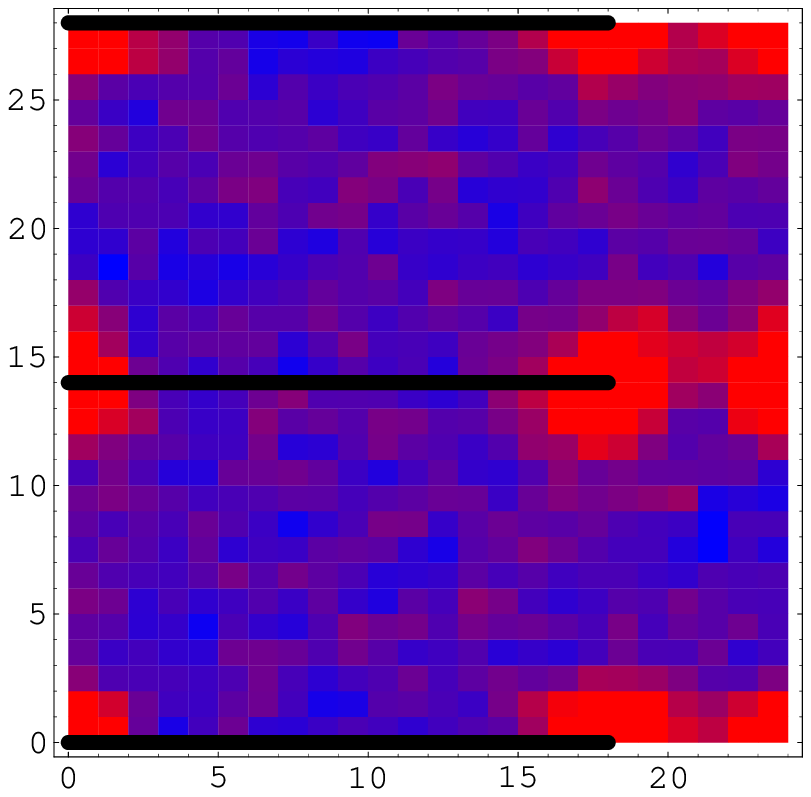}
  \includegraphics[width=6cm]{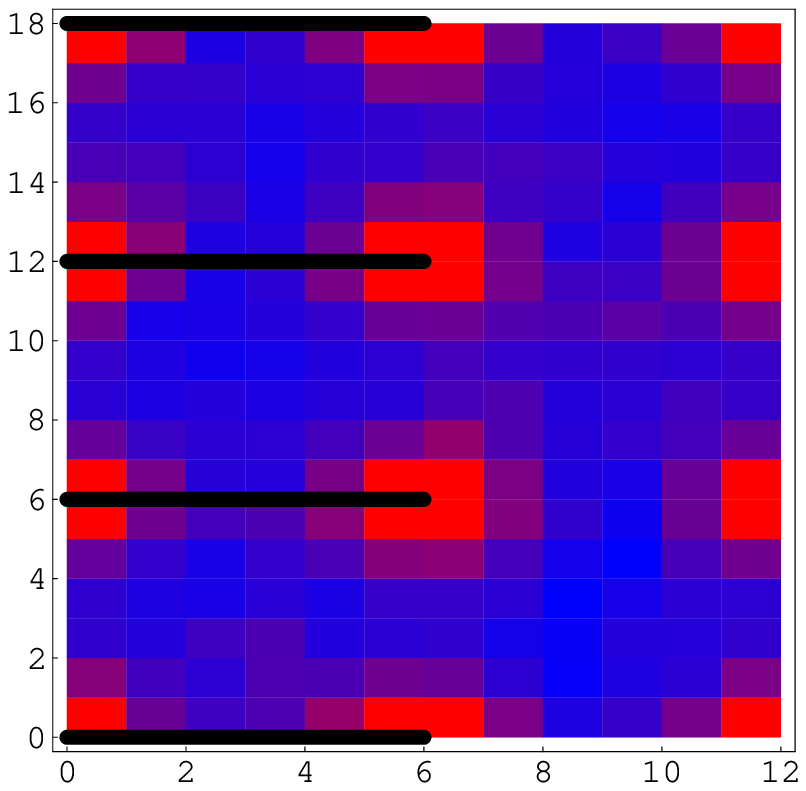}\\
  \caption{The distributions of average action density in the $\lr{x^{0},x^{1}}$ plane for different lattices and cut sizes:
 at the upper left: $24 \times 20^{3}$ lattice with two cuts, $a^{-1} l = 8$, $a = 0.12 \, fm$,
 at the upper right: $24 \times 20^{3}$ lattice with two cuts, $a^{-1} l = 6$, $a = 0.12 \, fm$,
 at the lower left: $28 \times 24^{3}$ lattice with two cuts, $a^{-1} l = 6$, $a = 0.10 \, fm$,
 at the lower right: $18 \times 12^{3}$ lattice with three cuts, $a^{-1} l = 6$, $a = 0.17 \, fm$.
 The cuts are shown as thick solid lines. The topology of lattices with two cuts is illustrated on Fig. \ref{fig:cutted_box}.}
  \label{fig:cut_g2}
\end{figure*}

\begin{figure}[ht]
  \includegraphics[width=6cm, angle=-90]{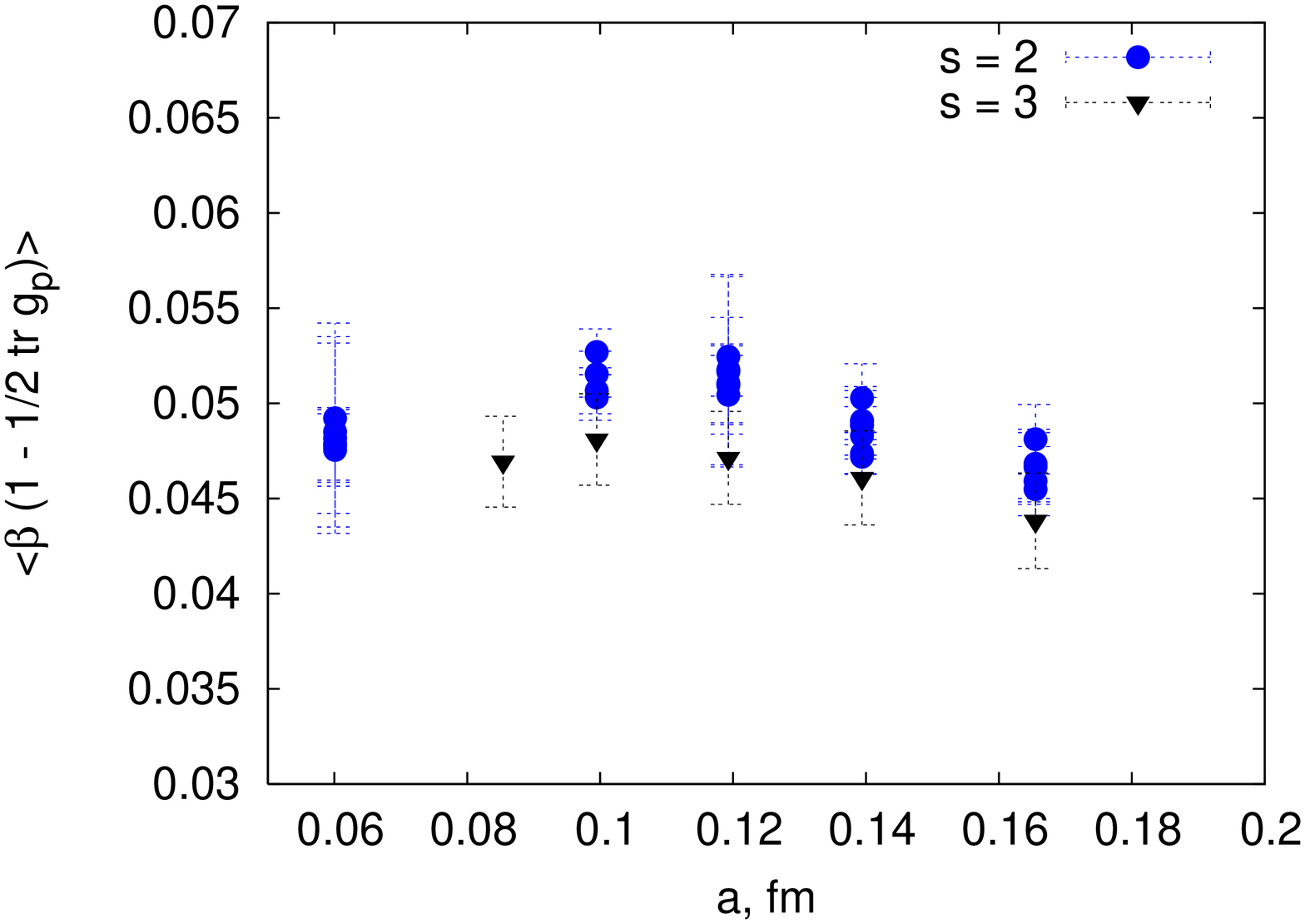}\\
  \caption{Average excess of action on the lattice plaquettes which are closest to the branching points on different lattices with two and three cuts.}
  \label{fig:DF_g23}
\end{figure}

 To check the physical scaling of the first, UV divergent term in (\ref{ent_asymp}), we have performed lattice Monte-Carlo simulations on lattices with cuts at different values of $l$ and various lattice spacings, lattice sizes and numbers of cuts. We have used from $30$ to $50$ lattice configurations obtained on lattices with sizes from $28 \times 24^{3}$ to $18 \times 12^{3}$ and with lattice spacings from $a = 0.06 \, fm$ to $a = 0.17 \, fm$. The number of cuts was equal to $2$ for all lattices except for the $18 \times 12^{3}$ lattice, which had three cuts. The action density for a given lattice site was obtained by averaging the action of all plaquettes to which this site belongs. The distributions of the average action density in the $\lr{x^{0},x^{1}}$ plane of the $s N_{t} \times N_{L}^{3}$ toric lattices which cover the original lattices with cuts are plotted on Fig. \ref{fig:cut_g2} for different lattice parameters. The cuts are shown as thick solid lines. Numerically the average plaquette action appears to be much higher in the immediate vicinity of the branching points, namely, for the lattice plaquettes which are closest to them. Since in four dimensions the branching points are situated on the two-dimensional boundary of $A$, one can expect that this excess of action near the branching points is related to the first, quadratically divergent term in the entanglement entropy (\ref{ent_asymp}). In order to check this, the average action density was measured for those sites on the $s N_{t} \times N_{L}^{3}$ toric covering lattice which are closest to each branching point (see Fig. \ref{fig:cut_g2}, where the action density is maximal in these points) for different lattices with two and three cuts. According to (\ref{ent_vs_fe}), (\ref{free_enrg_vs_plaq}) and (\ref{working_def}), the average action density for lattices without cuts was subtracted. The results of these measurements are plotted on Fig. \ref{fig:DF_g23} as the function of lattice spacing $a$. One can see that the average action excess practically does not depend on the lattice spacing, and that for two and three cuts the results are the same within the range of statistical errors. Since at fixed lattice size the branching points cover the surface with total area $s N_{L}^{2} a^{2}$ on the covering $s N_{t} \times N_{L}^{3}$ lattice, the total action excess associated with the boundary of $A$ includes the quadratically divergent term $\sim s a^{-2} |\partial A|$. According to (\ref{free_enrg_vs_plaq}), the same is true for $F\lrs{A, s, T}$. Differentiating over $s$, we thus recover the first, quadratically divergent term in (\ref{ent_asymp}). Since the divergence in the total average action comes from the vicinities of the branching points, it is clear that the divergent term in the free energy (\ref{free_enrg_vs_plaq}) and the entanglement entropy (\ref{ent_vs_fe}) does not depend on $l$.

\section{UV finite part of the entanglement entropy}
\label{sec:ent_growth}

 In this Section we present the results of our measurements of the derivative of the UV finite part of the entanglement entropy and point out a signature of its discontinuity.

 As discussed in Section \ref{sec:intro}, only the nondivergent part $S_{f}\lr{l}$ of the entanglement entropy depends on $l$. In our simulations we have found that it is very difficult to extract $S_{f}\lr{l}$ from the measurements of the average action excess using (\ref{free_enrg_vs_plaq}) and  integrating the average action over $\beta$ (see Section \ref{sec:ent_path_int}). To overcome this difficulty, we implement the method proposed in \cite{Fodor:07:1, Fodor:07:2} and directly measure the differences of the free energies $F\lr{l + a, 2, T} - F\lr{l,2,T}$ in (\ref{working_def}) at different $l$. Another advantage of this method is that all ultraviolet divergences are automatically canceled because the UV divergent part of the entanglement entropy does not depend on $l$, and no subtractions of $F\lr{T}$ should be made in (\ref{working_def}). The method itself is briefly described in Appendix \ref{sec:mc_fe}. All measurements were performed on the $24 \times 12^{3}$ lattice, and the distance $l$ was varied both by changing the lattice spacing $a$ at fixed $l/a$ and by changing $l/a$ at fixed $a$. Lattice spacings and the lengths $l$ used in our simulations are listed in Table \ref{tab:sim_par}, together with the number of lattice configurations used to obtain each data point. The method of \cite{Fodor:07:1, Fodor:07:2} involves integration of the outcome of Monte-Carlo simulations over a certain parameter $\alpha \in \lrs{0, 1}$ (see Appendix \ref{sec:mc_fe}). To perform this integration we have calculated the integrand at $\alpha = 0, \; 0.2, \; 0.4, \; 0.6, \; 0.8, \; 1.0$, and the numbers of lattice configurations in Table \ref{tab:sim_par} are the numbers of configurations for each of these values of $\alpha$.

\begin{table}
\centering
\begin{tabular}{|c|c|c|c||c|c|c|c|}
  \hline
  $\beta$ & $a, \; fm$ & $a^{-1} l$ & \# conf. & $\beta$ & $a, \; fm$ & $a^{-1} l$ & \# conf. \\
  \hline
  2.35  & 0.1394 & 2 & 38 & 2.425 & 0.1113 & 5 & 98  \\
  2.35  & 0.1394 & 4 & 19 & 2.45  & 0.0996 & 2 & 49  \\
  2.35  & 0.1394 & 5 & 97 &  2.45  & 0.0996 & 4 & 97  \\
  2.40  & 0.1193 & 3 & 98 & 2.45  & 0.0996 & 5 & 141 \\
  2.40  & 0.1193 & 4 & 98 & 2.50  & 0.0854 & 2 & 18  \\
  2.40  & 0.1193 & 5 & 98 &       &        &   &    \\
  \hline
\end{tabular}
\caption{The data points which were used to estimate the derivative $\frac{\partial}{\partial l} \, S_{f}\lr{l}$.}
\label{tab:sim_par}
\end{table}

\begin{figure}[ht]
  \includegraphics[width=6cm, angle=-90]{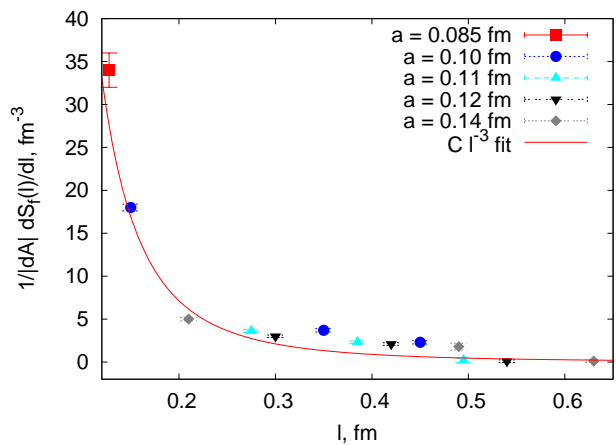}\\
  \caption{The dependence of the derivative of the entanglement entropy $\frac{1}{|\partial A|} \, \frac{S_{f}\lr{l}}{\partial l}$ on $l$. Solid line is the fit of the data by the function $C \, l^{-3}$.}
  \label{fig:Sphys_vs_lphys}
\end{figure}

 Finite differences $F\lr{l + a, 2, T} - F\lr{l,2,T}$ were then used to estimate the derivative $\frac{1}{|\partial A|} \, \frac{\partial}{\partial l} \, S_{f}\lr{l}$ in (\ref{ent_der_vs_fe_der}):
\begin{eqnarray}
\label{deriv_est}
\frac{1}{|\partial A|} \, \frac{\partial}{\partial l} \, S_{f}\lr{l + a/2} \approx
\frac{F\lr{l + a, 2, T} - F\lr{l,2,T}}{a \; |\partial A|}
\end{eqnarray}
The derivative $\frac{1}{|\partial A|} \, \frac{\partial}{\partial l} \, S_{f}\lr{l}$ estimated from these measurements is plotted on Fig. \ref{fig:Sphys_vs_lphys}. It can be seen that this derivative grows rapidly at small distances. For comparison with the asymptotic behavior $S_{f}\lr{l} \sim - l^{-2}$ at $l \rightarrow 0$, we have fitted these results by the function $C l^{-3}$ (solid line on Fig. \ref{fig:Sphys_vs_lphys}). For the data points with the smallest $l$ the finite differences (\ref{deriv_est}) were found at fixed $l/a$ at different $a$, so that the finite differences $\lr{l + a/2}^{-2} - \lr{l - a/2}^{-2}$ still behave as $l^{-3} \sim a^{-3}$. The estimated ratio of derivatives $\frac{\partial}{\partial l} \, S_{f}\lr{2 a_{2}} / \frac{\partial}{\partial l} \, S_{f}\lr{2 a_{1}} = \lr{2.0 \pm 0.4}$ at $a_{1} = 0.0996 \, fm$ and $a_{2} = 0.0854 \, fm$ is indeed close to the ratio $\lr{a_{1}/a_{2}}^{3} = 1.586$, which is an indication that at small $l$ the derivative $\frac{\partial}{\partial l} \, S_{f}\lr{l}$ indeed scales in physical units of length.

\begin{figure}[ht]
  \includegraphics[width=6cm, angle=-90]{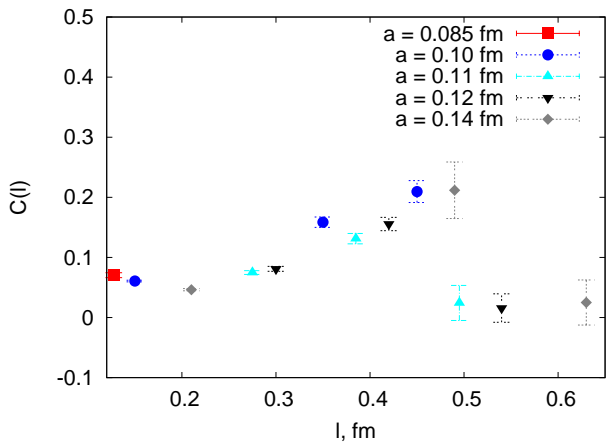}\\
  \caption{Entropic $C$-function $C\lr{l} = \frac{l^{3}}{|\partial A|} \, \frac{\partial}{\partial l} \, S\lr{l}$.}
  \label{fig:ecf}
\end{figure}

 At larger $l$ $\frac{\partial}{\partial l} \, S_{f}\lr{l}$ goes to zero faster than $l^{-3}$, and seem to approach a kind of plateau for the values of $l$ between $0.3 \, fm$ and $0.5 \, fm$. Here the values of $\frac{\partial}{\partial l} \, S_{f}\lr{l}$ obtained for different values of lattice spacing differ rather significantly, which indicates that for our lattice parameters finite-volume and finite-spacing effects may still be rather strong. Nevertheless, at least qualitatively all data points for different values of $a$ display the same behavior. We have also tried to plot the entropic $C$-function as the function of $l$ (see Fig. \ref{fig:ecf}). It also has a distinct discontinuity, however, it grows slowly for intermediate values of $l$ and thus its behavior seems to disagree with general theoretical expectations (see Section \ref{sec:intro}). In our opinion this growth is simply the artefact of the finite differences that we have used to estimate the derivative of the entanglement entropy over $l$. For example, if the finite difference $\frac{\Delta}{\Delta x} x^{-2}$ with $\Delta x = 1$ is used to estimate the function $ - x^{3} \, \frac{d}{d x} \, x^{-2}$, we obtain the function $x^{3} \, \frac{\Delta}{\Delta x} x^{-2} = x - \frac{x^{3}}{\lr{x+1}^{2}}$, which indeed grows slowly.

\begin{figure}[ht]
  \includegraphics[width=6cm, angle=-90]{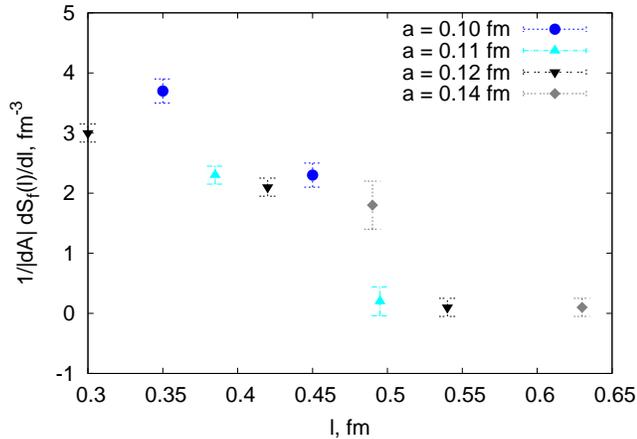}\\
  \caption{The discontinuity of the derivative of the entanglement entropy over $l$ near $l_{c} \approx 0.5 \, fm$. }
  \label{fig:Sphys_vs_lphys_PT}
\end{figure}

 What is most interesting, however, is that at $l_{c} \approx 0.5 \, fm$ the estimated derivative $\frac{\partial}{\partial l} \, S_{f}\lr{l}$ and the entropic $C$-function rapidly go to zero, and remain equal to zero within error range for larger values of $l$. Data points near this $l_{c}$ are plotted with larger scale on Fig. \ref{fig:Sphys_vs_lphys_PT}. This is a clear signature of the discontinuity of the derivative of the entanglement entropy over $l$, thus the entanglement entropy indeed seems to be nonanalytic in $l$ even at finite number of colours. Of course, it should be remembered that these results were obtained under some simplifying assumptions, in particular, the contributions of the free energies $F\lrs{A, s, T}$ with $s>2$ to the entanglement entropy (\ref{ent_vs_fe}) were neglected. One can not completely exclude the possibility that once higher $s$ are included in the analysis, the dependence of the entanglement entropy on $l$ may be changed. It could be therefore interesting to perform similar measurements on larger lattices with larger number of cuts.

\section{Confinement-deconfinement phase transition w.r.t. the size of $A$ at finite temperatures}
\label{sec:lt_phase_tr}

 If the expression (\ref{ent_vs_fe}) is applied to theories at finite temperatures, very straightforward arguments suggest that the nonanalytic behavior of the entanglement entropy with respect to the size of $A$ may persist even at temperatures above the temperature of the confinement-deconfinement phase transition. In this case it is also very simple to demonstrate the relation between the deconfinement phase transition at $T = T_{c}$ and the discontinuity of the derivative of the entanglement entropy over $l$. Indeed, consider gauge theory at some temperature $T > T_{c}$. It is clear that when $A$ becomes very large and extends through the whole physical space, the free energy $F\lrs{A, s, T}$ is equal to the free energy of the theory at the temperature $T/s$, and one can always find such $s$ that $T/s < T_{c}$ and the theory is in the confinement phase. On the other hand, if the size of $A$ tends to zero, $F\lrs{A, s, T}$ is just the free energy of $s$ copies of the gauge fields at the temperature $T$, each of which is in the deconfinement phase. Since for the confinement-deconfinement phase transition in pure $SU\lr{2}$ lattice gauge theory there is an exact order parameter associated with $Z_{2}$ center symmetry, namely, the Polyakov loop, one can expect that the transition between confinement and deconfinement phases should be associated with some discontinuity in $l$, so that the center symmetry is broken spontaneously for $l < l_{c}$. If this is true, at this transition the free energy $F\lrs{A, s, T}$ should change as at the deconfinement phase transition at $T = T_{c}$, which is of the second order for pure $SU\lr{2}$ Yang-Mills theory. Thus the second derivative of $F\lrs{A, s, T}$ over $l$ should change stepwise at some $l = l_{c}\lr{s}$, and for the values of $l$ close to $l_{c}$ we can write $F\lr{A, s, T} \sim \lr{l_{c}\lr{s} - l}^{2} \, \theta\lr{l_{c}\lr{s} - l} \; + \; \lr{ \, \ldots \,}$, where $\lr{\ldots}$ denotes the analytic part of $F\lr{A, s, T}$. Differentiating over $s$ and $l$, we find that if the limit $l_{c}^{\lr{0}} = \lim \limits_{s \rightarrow 1} l_{c}\lr{s}$ is different from zero, the derivative of the entanglement entropy over $l$ should have a discontinuity of the form $\frac{\partial}{\partial l}\, S_{f}\lrs{A} \sim \theta\lr{l_{c}^{\lr{0}} - l} \; \lim \limits_{s \rightarrow 1} \frac{\partial }{\partial s} \, l_{c}\lr{s}$. However, even if $\lim \limits_{s \rightarrow 1} l_{c}\lr{s} = 0$ and the entanglement entropy is analytic in $l$, one can still discuss the non-analytic behavior of the purity $P = \tr \hat{\rho}_{A}^{2}$ and its generalizations like $P_{A}\lr{s} = \tr_{A} \hat{\rho}_{A}^{s}$, which also indicates that near $l_{c}\lr{s}$ quantum entanglement between the regions $A$ and $B$ is quickly destroyed. In general all the quantities $P_{A}\lr{s}$ are nonanalytic in $l$ at different $l_{c}\lr{s}$, which is a typical situation when a phase transition turns into a crossover and different order parameters undergo transitions at different temperatures. It is also interesting to note that such transitions with respect to the size of $A$ can occur in any theories with phase transitions at finite temperatures.

\begin{figure}[ht]
  \includegraphics[width=6cm, angle=-90]{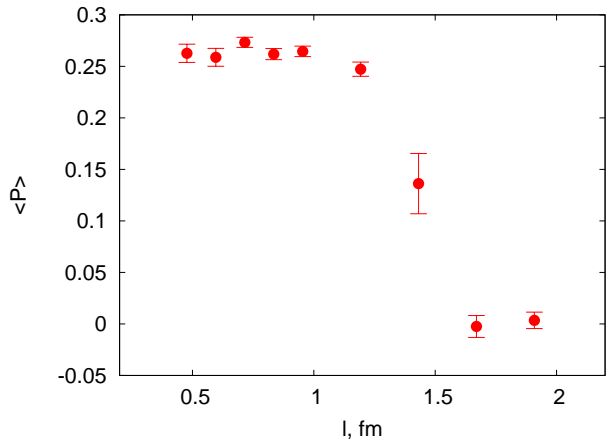}\\
  \caption{The expectation value of the Polyakov loop which winds around the "short" cycles of the lattice as the function of $l$ for $8 \times 20^{3}$ lattice with two cuts, $a = 0.12 fm$, $T/T_{c} = 1.43$.}
  \label{fig:pl_vs_l}
\end{figure}

 To study this transition from deconfinement to confinement, we have measured the expectation value of the Polyakov line which winds around the cycle of length $\lr{a T}^{-1} = 4$ on the $8 \times 20^{3}$ lattice with two cuts of variable length (see Fig. \ref{fig:cutted_box}) at $\beta = 2.40$. If the cut goes through the whole lattice, we have two lattices with $N_{t} = 4$, for which the confinement-deconfinement transition occurs near $\beta = 2.29$ \cite{Karsch:92:1}, therefore at $\beta = 2.40$ we are indeed in the deconfinement phase. On the other hand, with no cuts the time extent of the lattice is $N_{t} = 8$ and the deconfinement transition occurs at $\beta = 2.51$ \cite{Karsch:92:1}, thus we are in the confinement phase. Expectation value of the Polyakov loop as the function of $l$ is plotted on Fig. \ref{fig:pl_vs_l}. We see that at $l$ approximately equal to $1.5 \; fm$, the expectation value of the Polyakov loop has a sharp bend, after which it is equal to zero within error range. Thus center symmetry is restored and we are indeed in the confinement phase. Such dependence of the Polyakov loop on $l$ is qualitatively the same as its dependence on temperature at the usual confinement-deconfinement phase transition in pure $SU\lr{2}$ Yang-Mills theory, in agreement with the predictions of \cite{Takayanagi:06:1, Ryu:06:1, Ryu:06:2, Klebanov:07:1}.

\section{Conclusions}
\label{sec:conclusions}

 In this work we have studied numerically the behavior of the entropy of entanglement between a three-dimensional slab $A$ of thickness $l$ and its complement for $SU\lr{2}$ lattice gauge theory. This entropy is found to be a sum of the quadratically divergent and the UV finite terms $S_{UV}$ and $S_{f}\lr{l}$, both of which are proportional to the area of the boundary of $A$. The behavior of the nondivergent term was compared with the results of recent theoretical works \cite{Takayanagi:06:1, Klebanov:07:1, Velytsky:08:1}, where a discontinuity of the derivative of the entanglement entropy at some critical value $l = l_{c}$ was predicted. We have indeed found a signature of such discontinuity at $l_{c} \approx 0.5 \; fm$, where the entanglement entropy extracted from finite differences (\ref{deriv_est}) and (\ref{working_def}) rapidly goes to zero. Thus the predictions of \cite{Takayanagi:06:1, Klebanov:07:1} are confirmed for finite $N_{c}$, at least qualitatively. It seems that such behavior of entanglement entropy is a common feature of all confining gauge theories, similarly to the confinement-deconfinement phase transition at high temperatures. Since the entropic $C$-function $C\lr{l}$ measures the number of effective degrees of freedom at the energy scale $E \sim l^{-1}$, the results presented in \cite{Takayanagi:06:1, Klebanov:07:1} and in this paper imply that the transition from asymptotic freedom to confinement is always rapid, either when the temperature or simply the energy scale are varied. In other words, gluons and glueballs can not coexist at one energy scale, and the properties of gauge theories should be described either in terms of coloured or colourless particles, but never in terms of both.

\begin{figure}[ht]
  \includegraphics[width=7cm]{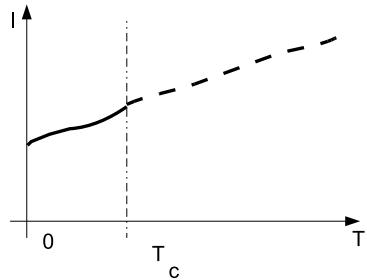}\\
  \caption{A conjectured line of phase transitions in the $l - T$ plane.}
  \label{fig:lt_phase_tr}
\end{figure}

 At finite temperatures above $T_{c}$ we have found that for lattices with cuts there is a real phase transition from deconfinement at small $l$ to confinement at large $l$. This transition is associated with spontaneous breaking of the $Z_{2}$ symmetry of the Polyakov loop, just as in the case of the usual confinement-deconfinement phase transition. As is argued in Section \ref{sec:lt_phase_tr}, such transition should also be associated with a non-analytic behavior of the purity $P = \tr \hat{\rho}_{A}^{2}$ of the reduced density matrix $\hat{\rho}_{A}$. More generally, for lattices with $s$ cuts at $T > T_{c}$ the deconfinement phase transition w.r.t. $l$ at some $l_{c}\lr{s}$ results in a non-analytic behavior of the quantity $P_{A}\lr{s} = \tr \hat{\rho}_{A}^{s}$. Since in general all $l_{c}\lr{s}$ are different, different measures of entanglement yield different transition points, which is a typical behavior for crossover. Thus it reasonable to conjecture that the line of phase transitions w.r.t. $l$ in the $l - T$ plane can be continued above $T_{c}$ as a sort of crossover line (see Fig. \ref{fig:lt_phase_tr}), which is associated with rapid but continuous destruction of quantum entanglement.

 It can be interesting to investigate how such finite-temperature phase transition for the entanglement entropy can be described in the framework of AdS/CFT correspondence, and how the geometry of the minimal hypersurface in the bulk which spans on the boundary of $A$ changes at this transition. On the other hand, a more detailed study of the line of confinement-deconfinement phase transitions in the $l - T$ plane in lattice or continuum gauge theories can also provide some information on higher-dimensional geometries which are dual to confining gauge theories. Finally, we would like to explain why, in our opinion, the confirmation of the predictions of \cite{Takayanagi:06:1, Ryu:06:1, Ryu:06:2, Klebanov:07:1} is important methodologically. Until recently the development of the AdS/QCD phenomenology was based on the attempts to fit some known data with the results of calculations of some geometric quantities in the dual higher-dimensional geometries. In contrast, the nonanalytic behavior of the entanglement entropy was not previously known for confining gauge theories and was originally found in \cite{Takayanagi:06:1} using the dual description, which demonstrates the predictive power of the AdS/QCD approach.

\begin{acknowledgments}
 The authors are grateful to V. I. Zakharov and T. Takayanagi for interesting discussions and for bringing the papers \cite{Klebanov:07:1} and \cite{Takayanagi:06:1} to our attention, and to E. T. Akhmedov and F. V. Gubarev for useful remarks. This work was partly supported by grants RFBR 06-02-04010-NNIO-a, RFBR 08-02-00661-a, DFG-RFBR 436 RUS, grant for scientific schools NSh-679.2008.2 and by Federal Program of the Russian Ministry of Industry, Science and Technology No 40.052.1.1.1112 and by Russian Federal Agency for Nuclear Power.
\end{acknowledgments}

\appendix

\section{Path integral representation of the entanglement entropy}
\label{sec:path_int_der}

 This Appendix revises the derivation of the path-integral representation for the entanglement entropy, which was originally obtained in \cite{Calabrese:04:1}. We first represent the entanglement entropy in the following form:
\begin{eqnarray}
\label{ent_ent_via_purity}
S_{A} = - \lim \limits_{s \rightarrow 1} \frac{\partial}{\partial s} \, \ln \tr_{A} \hat{\rho}_{A}^{s}
\end{eqnarray}
To obtain the reduced density matrix $\hat{\rho}_{A} = \tr_{B} \hat{\rho}_{AB} = \tr_{B} \ket{0} \bra{0}$ of the ground state of the theory, it is convenient to consider the theory at some finite temperature $T$, so that $\hat{\rho}_{AB} = \mathcal{Z}^{-1}\lr{T} \expa{ - T^{-1} \, \mathcal{H}}$, where $\mathcal{H}$ is the Hamiltonian of the theory and $\mathcal{Z}\lr{T} = \tr \expa{ - T^{-1} \, \mathcal{H}}$ is the partition function. At the end of the calculation one can take the limit $T \rightarrow 0$.

 We denote the set of all fields in the theory as $\phi\lr{\vec{x}}$, where $\vec{x}$ denotes spatial coordinates, omitting all indices for the sake of brevity. For gauge theories one should also include into $\phi$ the appropriate ghost sector. According to the Feynman-Kac path integral formula \cite{KapustaFiniteTemperatureQFT}, matrix elements of $\expa{ - T^{-1} \hat{\mathcal{H}} }$ between the eigenstates $\ket{\phi'\lr{\vec{x}}}$, $\ket{\phi''\lr{\vec{x}}}$ of the field operators are given by the path integral over all fields on a section of four-dimensional Euclidean space of time extent $T^{-1}$, with corresponding eigenvalues set as boundary conditions at $t = 0$ and $t = T^{-1}$ \cite{KapustaFiniteTemperatureQFT}:
\begin{widetext}
\begin{eqnarray}
\label{dm_pathint1}
\rho_{AB}\lrs{ \phi'\lr{\vec{x}}, \phi''\lr{\vec{x}} } =
\bra{\phi'\lr{\vec{x}}} \hat{\rho}_{AB} \ket{\phi''\lr{\vec{x}}}
= \nonumber \\ =
\mathcal{Z}^{-1}\lr{T} \,
\int \limits_{\phi\lr{\vec{x},0} =  \phi'\lr{\vec{x}} }^{ \phi\lr{\vec{x},T^{-1}} =  \phi''\lr{\vec{x}} }
\mathcal{D}\phi\lr{\vec{x},t}\;
\expa{ - \int \limits_{0}^{T^{-1}} dt \int d^{3}\vec{x} \mathcal{L}\lrs{\phi\lr{\vec{x},t}} }
\end{eqnarray}
where $\mathcal{L}\lrs{\phi\lr{\vec{x},t}}$ is the Lagrangian of the theory.

 In order to find the reduced density matrix for the region $A$, we should somehow trace over all degrees of freedom in the complement $B$ to the region $A$. Assume that the theory is regularized by discretizing the spatial coordinates $\vec{x}$. In this case it is always possible to rewrite the states $\ket{\phi\lr{\vec{x}}}$ as a direct product $\prod \limits_{\vec{x}} \ket{\phi_{\vec{x}}}$, where $\ket{\phi_{\vec{x}}}$ are the eigenstates of the field operator in the point $\vec{x}$. It is then natural to define the trace over all degrees of freedom in $B$ as as a sum over all products of basis states $\ket{\phi_{\vec{x}}}$ with $\vec{x} \in B$, so that the reduced density matrix $\hat{\rho}_{A}$ reads:
\begin{eqnarray}
\label{dm_pathint2}
\rho_{A}\lrs{ \phi'\lr{\vec{x}}, \phi''\lr{\vec{x}} } =
\int \limits_{\vec{x} \in B} \mathcal{D}\phi\lr{\vec{x}} \,
\bra{\phi\lr{\vec{x}}} \bra{\phi'\lr{\vec{x}}} \hat{\rho}_{AB}
 \ket{\phi''\lr{\vec{x}}} \ket{\phi\lr{\vec{x}}}
= \nonumber \\ =
\mathcal{Z}^{-1}\lr{T} \, \int \mathcal{D}\phi\lr{\vec{x},t}\;
\expa{ - \int \limits_{0}^{T^{-1}} dt \int d^{3}\vec{x} \mathcal{L}\lrs{\phi\lr{\vec{x},t}} }
\end{eqnarray}
where in the last path integral the following boundary conditions are imposed: $\phi\lr{\vec{x},0} = \phi'\lr{\vec{x}}$, $\phi\lr{\vec{x},T^{-1}} = \phi''\lr{\vec{x}}$ for $x \in A$, and $\phi\lr{\vec{x},0} = \phi\lr{\vec{x},T^{-1}}$, if $x \in B$. Matrix multiplication can also be defined in terms of the path integral over $\phi\lr{\vec{x}}$:
\begin{eqnarray}
\label{matrix_multipl_pathint}
\rho_{A}^{2}\lrs{ \phi'\lr{\vec{x}}, \phi''\lr{\vec{x}} }
=
\int \limits_{\vec{x} \in A} \mathcal{D}\phi\lr{\vec{x}} \: \rho_{A}\lrs{ \phi'\lr{\vec{x}}, \phi\lr{\vec{x}} }
\rho_{A}\lrs{ \phi\lr{\vec{x}}, \phi''\lr{\vec{x}} }
\end{eqnarray}
 It is now straightforward to calculate the trace and the matrix contractions in the definition of the function $P_{A}\lr{s} = \tr \hat{\rho}_{A}^{s}$:
\begin{eqnarray}
\label{dm_pathint3}
\tr \hat{\rho}_{A}^{s} =
\int \limits_{\vec{x} \in A}
\mathcal{D}\phi_{1}\lr{\vec{x}}
\ldots
\mathcal{D}\phi_{s}\lr{\vec{x}} \nonumber \\
\rho_{A}\lrs{ \phi_{1}\lr{\vec{x}}, \phi_{2}\lr{\vec{x}}  }
\rho_{A}\lrs{ \phi_{2}\lr{\vec{x}}, \phi_{3}\lr{\vec{x}}  }
\ldots
\rho_{A}\lrs{ \phi_{s}\lr{\vec{x}}, \phi_{1}\lr{\vec{x}}  }
\end{eqnarray}
\end{widetext}
Putting together the expressions (\ref{dm_pathint2}) and (\ref{dm_pathint3}), one sees that to calculate $\tr \hat{\rho}_{A}^{s}$ in (\ref{dm_pathint3}), one should impose periodic boundary conditions with period $s T^{-1}$ in time if $\vec{x} \in A$ and with period $T^{-1}$ if $\vec{x} \in B$, and find the partition function by integrating over all such fields. We thus arrive at the following representation for $\tr \hat{\rho}_{A}^{s}$:
\begin{eqnarray}
\label{purity_via_pf}
\tr \hat{\rho}_{A}^{s} =
\frac{\mathcal{Z}\lrs{A, s, T}}
{\mathcal{Z}^{s}\lr{T}}
\end{eqnarray}
where $\mathcal{Z}\lrs{A, s, T}$ is the partition function obtained by integrating over all fields which satisfy the $\vec{x}$-dependent periodic boundary conditions, as discussed above. The topology of space on which the fields which enter the path integral for $\mathcal{Z}\lrs{A, s ,T}$ live is schematically shown on Fig. \ref{fig:cutted_box} for $s = 2$.

 Inserting the expression (\ref{purity_via_pf}) into (\ref{ent_ent_via_purity}) and taking the limit $T \rightarrow 0$, we arrive at the following representation of the entanglement entropy:
\begin{eqnarray}
\label{ent_vs_fe1}
S\lrs{A} =
\lim \limits_{T \rightarrow 0}
\lr{
\lim \limits_{s \rightarrow 1} \, \frac{\partial}{\partial s} \, F\lrs{A, s, T}
- F\lr{T}
}
\end{eqnarray}
where $F\lr{T} = - \ln \mathcal{Z}\lr{T}$ and $F\lrs{A, s, T} = - \ln \mathcal{Z}\lrs{A, s, T}$ are the free energies which correspond to the partition functions $\mathcal{Z}\lr{T}$ and $\mathcal{Z}\lrs{A, s, T}$.

\section{Direct Monte-Carlo measurements of the differences of free energies}
\label{sec:mc_fe}

 In this Appendix we briefly review the method proposed in \cite{Fodor:07:1, Fodor:07:2} for direct measurements of the differences of free energies in lattice Monte-Carlo simulations. The works \cite{Fodor:07:1, Fodor:07:2} deal mainly with the case of finite-temperature lattice gauge theories, so it may be useful to repeat their arguments in a somewhat more general context.

 Assume that we have two different actions $S_{1}\lrs{\phi}$ and $S_{2}\lrs{\phi}$ for some set of fields $\phi$ defined on the lattice of some fixed size. Again, here $\phi$ denotes any lattice fields, which can live on lattice sites, links, cubes, etc., as long as the number of these lattice simplices which enter both actions remains the same. For instance, this is true for gauge fields on the lattices of special topology, which are described in Section \ref{sec:act_dist}.

\begin{figure}[ht]
  \includegraphics[width=6cm, angle=-90]{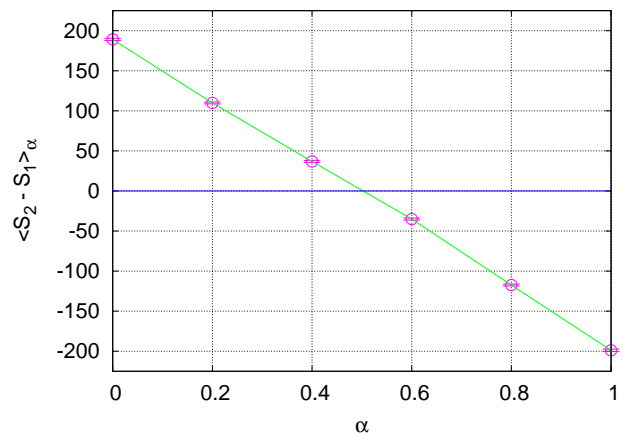}\\
  \caption{The dependence of the integrand in (\ref{pf_ratio_final}) on $\alpha$ for the difference of free energies on $24 \times 12^{3}$ lattices with two cuts of length $10 a$ and $11 a$, lattice spacing $a = 0.085 \, fm$.}
  \label{fig:integrand_vs_alpha250}
\end{figure}

 We would like to calculate the difference $F_{2} - F_{1}$ of the free energies $F_{1} = - \ln \mathcal{Z}_{1}$ and $F_{2} = - \ln \mathcal{Z}_{2}$, where $\mathcal{Z}_{1}$ and $\mathcal{Z}_{2}$ are the partition functions calculated by integrating over all fields $\phi$ with the weights $\expa{ - S_{1}\lrs{\phi}}$ and $\expa{ - S_{2}\lrs{\phi}}$, respectively. To this end, define the interpolating partition function $\mathcal{Z}\lr{\alpha}$, so that $\mathcal{Z}\lr{0} = \mathcal{Z}_{1}$ and $\mathcal{Z}\lr{1} = \mathcal{Z}_{2}$:
\begin{eqnarray}
\label{interpolating_pf}
\mathcal{Z}\lr{\alpha} = \int \mathcal{D}\phi \expa{
 - \lr{1 - \alpha} \, S_{1}\lrs{\phi}
 - \alpha \, S_{2}\lrs{\phi}
}
\end{eqnarray}
Now the difference $F_{2} - F_{1} = \ln \mathcal{Z}_{1} - \ln \mathcal{Z}_{2}$ can be represented in the following form:
\begin{eqnarray}
\label{pf_ratio1}
F_{2} - F_{1} = - \int \limits_{0}^{1} d\alpha \, \frac{\partial}{\partial \alpha} \,
\ln \mathcal{Z}\lr{\alpha}
\end{eqnarray}
Calculating the derivative over $\alpha$ from (\ref{interpolating_pf}), we arrive at the following result:
\begin{eqnarray}
\label{pf_ratio_final}
F_{2} - F_{1} = \int \limits_{0}^{1} d\alpha \,
\vev{
S_{2}\lrs{\phi} - S_{1}\lrs{\phi}
}_{\alpha}
\end{eqnarray}
where the average $\vev{\ldots}_{\alpha}$ is defined by integrating over all fields with the weight $\mathcal{Z}^{-1}\lr{\alpha} \; \expa{ - \lr{1 - \alpha} \, S_{1}\lrs{\phi} - \alpha \, S_{2}\lrs{\phi}}$. For any $\alpha \in \lrs{0,1}$, the average $\vev{S_{2}\lrs{\phi} - S_{1}\lrs{\phi}}_{\alpha}$ can be calculated with arbitrary precision by increasing the number of Monte-Carlo iterations. After that some numerical integration method should be used to find the integral in (\ref{pf_ratio_final}). In practice it turns out that although for each $\alpha$ the expectation value $\vev{S_{2}\lrs{\phi} - S_{1}\lrs{\phi}}_{\alpha}$ is numerically rather large, the contributions from $\alpha < 1/2$ and $\alpha > 1/2$ cancel almost exactly. For example, the dependence of the integrand in (\ref{pf_ratio_final}) on $\alpha$ for the difference of free energies on $24 \times 12^{3}$ lattices with two cuts with $l = 2 a$ and $l = 1 a$ is plotted on Fig. \ref{fig:integrand_vs_alpha250}. Thus the expectation value $\vev{S_{2}\lrs{\phi} - S_{1}\lrs{\phi}}_{\alpha}$ should be found with a very good precision, which typically requires rather large number of Monte-Carlo iterations. However, as discussed in the body of the text, in our case the described method seems to be the best way to measure free energies.

\fussy

\end{document}